\documentclass[10pt, conference, letterpaper]{IEEEtran}
\usepackage{times,amsmath,epsfig,verbatim}
\usepackage{subfigure,algorithm}
\usepackage{graphicx}
\usepackage{amssymb}
\usepackage{url}
\usepackage{color}
\usepackage{mathrsfs}
\usepackage{amsthm,amsfonts}
\usepackage{float}
\usepackage[square, comma, sort&compress, numbers]{natbib}
\usepackage{bm}
\usepackage{algorithmic}
\usepackage{bbm}

\graphicspath{{figs/}}

\newtheorem{corollary}{{\bf Corollary}}

\begin{document}

\title{Sampling Online Social Networks \\ via Heterogeneous Statistics}

\author{Xin Wang$^\dag$, Richard T. B. Ma$^*$, Yinlong Xu$^\dag$, Zhipeng Li$^\dag$ \\
$^\dag$ School of Computer Science and Technology, University of Science and Technology of China\\
$^*$ School of Computing, National University of Singapore\\
{\it \{yixinxa, lizhip\}@mail.ustc.edu.cn, tbma@comp.nus.edu.sg, ylxu@ustc.edu.cn}
}

\maketitle
\begin{abstract}
Most sampling techniques for online social networks (OSNs) are based on a particular sampling method on a single graph, which is referred to as a statistic. However, various realizing methods on different graphs could possibly be used in the same OSN, and they may lead to different sampling efficiencies, i.e., asymptotic variances.
To utilize multiple statistics for accurate measurements, we formulate a mixture sampling problem, through which we construct a mixture unbiased estimator which minimizes the asymptotic variance.
Given fixed sampling budgets for different statistics, we derive the optimal weights to combine the individual estimators; given a fixed total budget, we show that a greedy allocation towards the most efficient statistic is optimal.
In practice, the sampling efficiencies of statistics can be quite different for various targets and are unknown before sampling. To solve this problem, we design a two-stage framework which adaptively spends a partial budget to test different statistics and allocates the remaining budget to the inferred best statistic.
We show that our two-stage framework is a generalization of 1) randomly choosing a statistic and 2) evenly allocating the total budget among all available statistics, and our adaptive algorithm achieves higher efficiency than these benchmark strategies in theory and experiment.
\end{abstract}
\section{Introduction}
\setlength{\parindent}{1em}

With the ever increasing popularity of online social networks (OSNs) in recent years, many studies have focused on the analysis of OSNs, such as estimating various properties of the users and their relationships. 
OSNs are usually measured via graph sampling techniques, because they are typically too large to be completely visited and OSN service providers rarely make their complete network dataset publicly visible.
To guarantee the estimation accuracy, many unbiased graph sampling methods have been designed, such as the simple random walk with re-weighting (RWRW) \cite{A.H.Rasti2009Res, M.Gjoka2011Practical}, the frontier sampling (FS) \cite{ribeiro2010estimating} and the random walk with uniform restarts (RWuR) \cite{KAWAW}.
However, OSNs often consist of multiple social graphs which can be sampled by different unbiased graph sampling methods.
For example, in the YouTube social network, users are allowed to declare friendship with each other and create interest groups for others to join in. This creates two graphs whose edge sets correspond to 1) the mutual friendship and 2) the sharing of membership of some interest group among the users, respectively.
For a given measurement target, sampling via different graphs usually have different efficiencies, which also vary as the measurement target changes.
Furthermore, various graph sampling methods can be applied to the same social graph, e.g., the FS and the RWuR are both realizable in the friendship graph of LiveJournal. However, they might induce different sampling efficiencies, which are often unknown a priori. Although one can use multiple unbiased \emph{statistics}, generated by different methods on different graphs, to form a heterogeneous statistic,
it is unclear {\em how one could  1) optimally allocate the sampling budgets among different statistics and 2) optimally combine them. }


As we focus on unbiased estimators, we use the asymptotic variance \cite{Anton2001Ordering} to measure the efficiency of a statistic (or its estimator). We formulate a \emph{mixture sampling problem} that tries to minimize the asymptotic variance of a linearly mixed estimator, constrained by sampling budgets.
Given allocated budgets for different statistics, we prove that the optimal weights of individual estimators are inversely proportional to their asymptotic variances; under a fixed total budget, we rank the allocation decisions and find that a greedy allocation is optimal, i.e., allocating more budgets to the statistic with smaller asymptotic variance is always better.

However, the asymptotic variances of the statistics are usually unknown before sampling.
To address this challenge, we design a two-stage framework with a pilot and a regular sampling stage.
In the pilot sampling stage, we allocate part of the sampling budget to all the statistics and infer the most efficient statistic by estimating the asymptotic variance of each statistic.
In the regular sampling stage, we allocate the remaining budget to the inferred most efficient statistic.
Our framework is a generalization of two benchmark strategies: 1) spending all budget on a randomly chosen statistic and 2) allocating the budget among all available statistics evenly.
We show that our two-stage strategies achieve higher sampling efficiency than the two benchmark strategies. 
Furthermore, to allocate an optimal sub-budget for the pilot sampling stage, we design an online algorithm to dynamically estimate an upper-bound of the optimal fraction during the pilot sampling.
Because the inference of the most efficient statistic is made by estimating the asymptotic variances in the pilot sampling stage, it makes our framework \emph{adaptive} for different measurement targets.
Our framework does not restrict how the estimators of asymptotic variances should be constructed, as long as they are asymptotically unbiased.
To illustrate, we provide a detailed implementation and evaluate the performance of our framework in the Douban social network.
The experimental results show that our technique uses only $18\%-57\%$ of the sampling budget needed by the benchmark strategies for achieving the same estimation accuracy for a range of measurement targets.
Our main contributions are as follows.
\begin{itemize}
\item We formulate and solve a mixture sampling problem which constructs an optimal estimator of a heterogeneous statistic to improve sampling efficiency. In particular, we derive the optimal weights of the individual estimators in the mixture estimator  (Theorem \ref{the:basic1}) and the optimal allocation decisions among the statistics (Theorem \ref{the:basic2}).

\item We design a two-stage framework and an adaptive algorithm (Algorithm 1) for the pilot sampling, a practical solution for the mixture sampling problem when the efficiencies of the statistics are unknown before sampling.

\item We show that the two-stage strategies are asymptotically optimal (Theorem \ref{the:cond}) and achieve higher efficiency than two benchmark strategies (Corollary \ref{cor:1}).

\item As a case study, we provide a detailed implementation of our framework and evaluate its performance in the Douban social network.
\end{itemize}

The remaining of this paper is organized as follows.
Section \ref{sec:background} introduces the concepts and characteristics of unbiased graph sampling methods.
Section \ref{sec:pro} defines the mixture sampling problem and presents its optimal solution.
With unknown efficiencies of the statistics before sampling, we design the two-stage framework and its adaptive algorithm in Section \ref{sec:pilot}.
Section \ref{sec:sim} implements the framework and evaluates its performance in the Douban social network.
Section \ref{sec:related work} reviews related work and Section \ref{sec:conclusion} concludes.

\section{Unbiased Graph Sampling}\label{sec:background}
We denote an undirected graph in an online social network as $G \!=\! ({\cal V},{\cal E})$ with a set of nodes ${\cal V} \!= \!\{1,\cdots, V\}$ to represent users and a set of edges $\cal E$ to represent the relationships among the users. We denote $f$ as a property and $f_v$ as its value of user $v$. Our measurement target is to estimate the mean value of property $f$ over all users in $\cal V$, i.e., $\bar{f}\triangleq \left(\sum_{v\in {\cal V}}f_v\right)/V$.

We consider a graph sampling method that traverses the nodes of the graph via a random walk, which generates a discrete-time stochastic process $\{X_t\}_{t \in \mathbb{N}}$ with the state space of $\cal V$, i.e., $X_t \in \cal V$ for all $t\in \mathbb{N}$.
We define the random variable $\hat{f}(m)$ as an estimator on the sample path $\{X_t\colon t\!=\!1,\!\cdots\!,m\}$ of $m$ samples. An estimator $\hat{f}(\cdot)$ is {\em unbiased} if $E[\hat{f}(m)] = \bar f$ for all $m\in\mathbb{N}$ and is {\em asymptotically unbiased} if
{\setlength\abovedisplayskip{1.5pt}
\setlength\belowdisplayskip{1.5pt}
\begin{equation*}
\displaystyle \hat{f}(m) \xrightarrow{a.s.} \bar f \quad \text{as} \quad m\rightarrow\infty,
\end{equation*}
}
where $\xrightarrow{a.s.}$ denotes {\em convergence almost surely}. If the process $\{X_t\}_{t \in \mathbb{N}}$ is ergodic, by the central limit theorem (CLT), 
{\setlength\abovedisplayskip{1.5pt}
\setlength\belowdisplayskip{1.5pt}
\begin{equation}\label{eq:central}
\sqrt{m}[\hat{f}(m) - \bar{f}]\xrightarrow{d} N(0,\sigma^{2} (f)) \quad \text{as} \quad m\rightarrow\infty,
\end{equation}
where $\xrightarrow{d}$ denotes {\em convergence in distribution} and $N(0,\sigma^{2} (f))$ denotes a normal distribution with mean $0$ and variance $\sigma^{2}(f)$, which is defined by
\begin{equation}
\sigma^2 (f) \triangleq \displaystyle\lim_{m \rightarrow \infty} m Var(\hat{f}(m)). \label{eq:asy}
\end{equation}
}
By (\ref{eq:central}), we can infer that $\hat{f}(m) \!\xrightarrow{a.s.}\! \bar{f}$ as $m \!\rightarrow \!\infty$, i.e., $\hat{f}(m)$ is an asymptotically unbiased estimator of $\bar{f}$.
It also shows that the distribution of $\sqrt{m}\hat{f}(m)$ is asymptotically normal with variance $\sigma^2 (f)$,
which approximately determines how many samples are required to achieve a certain level of accuracy for the estimator $\hat{f}(m)$.
Thus, we use the asymptotic variance $\sigma^2 (f)$ to measure the efficiency of an asymptotically unbiased graph sampling method (or its estimator) in this paper.

In the next two sections, we formulate and solve a mixture sampling problem, based on which we design a two-stage framework to sample via multiple statistics. The estimators of these statistics can be based on very different asymptotically unbiased sampling methods on different graphs.
\section{Mixture Sampling Problem}
\label{sec:pro}

We consider an objective of measuring the mean value of property $f$ over the users, i.e., $\bar f$ defined earlier.
We refer to an asymptotically unbiased sampling method on a social graph
as a statistic, and assume there are $K$ types of statistics that can be applied in the OSN. For any statistic $k$, we denote the random variable $\hat{f}_k(m_k)$ as the value of its estimator given $m_k$ samples and $\sigma^2_k (f)$ as its asymptotic variance.
We simplify the notation $\sigma^2_k (f)$ as $\sigma_k^2$ when we focus on a single property $f$.
Because each estimator $\hat{f}_k(m_k)$ is asymptotically unbiased, we use the asymptotic variance $\sigma_k^2$ as a metric for comparing the efficiencies of these statistics. If the asymptotic variance $\sigma_i^2$ is smaller than $\sigma_j^2$
, we say statistic $i$ is more efficient than statistic $j$ for estimating $\bar{f}$. Furthermore, we denote $k^*$ as the most efficient statistic, i.e., $\sigma_{k^*}^2 = \min\{\sigma_k^2:k=1,\cdots,K\}$.

\subsection{Mixture Sampling Problem}
\label{sec:pro des}
Suppose we have a total sampling budget\footnote{We assume that one unit of the budget is the cost of visiting a node.} of $M$ samples and $K$ types of candidate statistics, we consider the mixture sampling problem of how to allocate the sampling budget among different statistics and how to construct an unbiased estimator $\hat{f}$ for $\bar f$ so as to minimize its asymptotic variance.

We denote $\bm{a} =(a_1,\cdots,a_K)$ as a budget allocation decision, where each $a_k\geq 0$ defines the fraction of the total budget allocated to statistic $k$. We define ${\cal K}_a \triangleq \{k: a_k > 0\}$ to be the set of active statistics. Thus, each active statistic $k$ has a budget $m_k = a_kM$ and an estimator $\hat{f}_k(m_k)$.
Because the sum of budget allocated to each statistic cannot exceed the total budget, we define the constraint set of the allocation decisions as ${\cal A} \triangleq \{\bm{a} | \sum_{k=1}^{K} a_k \leq 1; a_k\geq 0 \ \forall k=1,\cdots,K\}$.
Given a vector $\bm{\hat f} = ({\hat f}_1,\cdots,{\hat f}_K)$ of estimators, we consider a mixed estimator ${\hat f}(\bm{w})$ which linearly combines the individual estimators by a weight vector $\bm{w} =(w_1,\cdots,w_K)$, defined as
{\setlength\abovedisplayskip{1.5pt}
\setlength\belowdisplayskip{1.5pt}
\begin{equation}\label{eq:mixed}
\hat{f}(\bm{w})\triangleq \displaystyle\sum_{k=1}^K w_k \hat{f}_k.
\end{equation}
}
Each weight $w_k$ is used to determine the relative importance of the individual estimator $\hat{f}_k$.
Under a total budget $M$ and an allocation decision $\bm{a}$, we define the \emph{mixture estimator} with weights $\bm{w}$ as
{\setlength\abovedisplayskip{1.5pt}
\setlength\belowdisplayskip{1.5pt}
\begin{equation}\label{eq:ap_1}
\hat{f}(\bm{a},M,\bm{w})\triangleq  \displaystyle\sum_{k\in {\cal K}_a} w_k \cdot \hat{f}_k (m_k) = \displaystyle\sum_{k\in {\cal K}_a} w_k \cdot \hat{f}_k (a_kM).
\end{equation}
}
We define the asymptotic variance of the above estimator as
{\setlength\abovedisplayskip{1.5pt}
\setlength\belowdisplayskip{1.5pt}
\begin{equation}\label{eq:14asy}
\varsigma(\bm{a},\bm{w}) \triangleq \displaystyle\lim_{M \rightarrow \infty} M  \cdot Var(\hat{f}(\bm{a},M,\bm{w})).
\end{equation}
}
If each $\hat{f}_k$ is asymptotically unbiased, we hope that the constructed mixture estimator
$\hat{f}(\bm{a},M,\bm{w})$ would still be asymptotically unbiased. We denote the set ${\cal W}_{\bm{a}}$ to be the domain of weights under the budget allocation $\bm{a}$ such that for every $\bm{w}\in{\cal W}_{\bm{a}}$, $\hat{f}(\bm{a},M,\bm{w})$ is asymptotically unbiased.

Our design goal is to construct the optimal unbiased estimator $\hat{f}(\bm{a},M,\bm{w})$ whose asymptotic variance $\varsigma(\bm{a},\bm{w})$ could be minimized. We formulate two related {\em mixture sampling problems} as follows.
In the first problem, we consider a given allocation decision $\bm{a}$ and we denote $\varsigma_{\bm{a}}(\bm{w}) \triangleq \varsigma({\bm{a}},\bm{w})$. The objective is to find the optimal weights $\bm{w}^*$ that solve:
{\setlength\abovedisplayskip{1.5pt}
\setlength\belowdisplayskip{1.5pt}
\begin{equation}\label{eq:op1}
\displaystyle \text{Minimize} \quad \displaystyle \varsigma_{\bm{a}}(\bm{w}) \quad\quad  \text{subject to} \quad \displaystyle \bm{w}\in {\cal W}_{\bm{a}}.
\end{equation}
}
In the second problem, the objective is to find the optimal allocation decision $\bm{a}^*$ and the corresponding optimal weights $\bm{w}^*(\bm{a}^*)$ that solve:
{\setlength\abovedisplayskip{1.5pt}
\setlength\belowdisplayskip{1.5pt}
\begin{equation}\label{eq:op2}
\begin{aligned}
\displaystyle \text{Minimize}  & \quad \quad \displaystyle \varsigma(\bm{a},\bm{w}) \\
\quad\quad  \text{subject to}  & \quad \quad \displaystyle \bm{a}\in {\cal A} \quad \text{and} \quad \bm{w}\in {\cal W}_{\bm{a}}.
\end{aligned}
\end{equation}
}
The first problem can be regarded as a sub-problem of the second one, where the allocated decision is predetermined.

\subsection{Optimal Weights and Allocation Decisions}\label{sec:kav}

In this subsection, we solve the optimal weights to construct an estimator and the optimal budget allocation for maximizing
the efficiency of an estimator. Under a fixed budget allocation decision $\bm{a}$, intuitively, a larger weight $w_k$ should be given to an estimator
$\hat{f}_k$ if statistic $k$ is more efficient, i.e., its asymptotic variance $\sigma^2_k$ is smaller.
The following result provides an affirmative answer to the intuition.
\newtheorem{theorem}{Theorem}
\begin{theorem}\label{the:basic1}
Assume all the pure estimators $\hat{f}_k$ are independent of each other. The mixture estimator $\hat{f}(\bm{a},M,\bm{w})$ is asymptotically unbiased for $\bar{f}$ if and only if the domain of weights under an allocation decision $\bm a$ satisfies
{\setlength\abovedisplayskip{1.5pt}
\setlength\belowdisplayskip{1.5pt}
\begin{equation}\label{eq:ap_2}
{\cal W}_{\bm{a}} = \left\{\bm{w} | \sum_{k\in {\cal K}_{\bm a}} w_k = 1\right\}.
\end{equation}
}
Its asymptotic variance can be characterized by a function of the allocation $\bm{a}$ and the weight vector $\bm{w}$, defined as
{\setlength\abovedisplayskip{1.5pt}
\setlength\belowdisplayskip{1.5pt}
\begin{equation}\label{eq:ap_3}
\varsigma(\bm{a},\bm{w}) = \displaystyle\sum_{k\in {\cal K}_a} \!\frac{w_k^2}{a_k} \cdot\sigma_k^2.
\end{equation}
}
The optimal solution $\bm{w}^*$ of the optimization problem in Equation (\ref{eq:op1}) satisfies
{\setlength\abovedisplayskip{1.5pt}
\setlength\belowdisplayskip{1.5pt}
\begin{equation}\label{eq:optiwei}
w_k^* = \frac{a_k}{\sigma_k^{2}}/\displaystyle\sum_{i\in {\cal K}_a} \frac{a_i}{\sigma_i^{2}}, \quad \forall k\in {\cal K}_a,
\end{equation}
}
and the corresponding minimum asymptotic variance is
{\setlength\abovedisplayskip{1.5pt}
\setlength\belowdisplayskip{1.5pt}
\begin{equation*}
\varsigma_{\bm{a}}(\bm{w}^*)=\ \left[\displaystyle\sum_{k \in {\cal K}_a} \frac{a_k}{\sigma_k^{2}} \right]^{-1}.
\end{equation*}
}
\end{theorem}
Theorem \ref{the:basic1} shows that to guarantee the mixture estimator to be asymptotically unbiased, the sum of weights of the active statistics must be one. It also tells that when the allocation decision $\bm{a}$ is fixed, the optimal weight $w_k^*$ of each estimator $\hat{f}_k (m_k)$ is proportional to $a_k$ and inversely proportional to its asymptotic variance $\sigma_k^{2}$. Based on Theorem \ref{the:basic1}, we denote $\bm{w}^*(\bm{a})$ to be the optimal solution of (\ref{eq:op1}) defined in (\ref{eq:optiwei}) and the second optimization problem (\ref{eq:op2}) could be stated as finding the optimal allocation $\bm{a}^*$ that solves:
{\setlength\abovedisplayskip{1.5pt}
\setlength\belowdisplayskip{1.5pt}
\begin{equation}\label{eq:op3}
\displaystyle \text{Minimize}  \quad \displaystyle \varsigma(\bm{a},\bm{w}^*(\bm{a})) \quad  \text{subject to}  \quad \displaystyle \bm{a}\in {\cal A}.
\end{equation}
}
Intuitively, an optimal solution should allocate more budgets to the more efficient statistic. The next result shows that a greedy strategy that allocates all budgets to the statistic with the smallest asymptotic variance is actually optimal.

\begin{theorem}\label{the:basic2}
Assume that the conditions of Theorem \ref{the:basic1} hold. Denote $\{\sigma_{(k)}^2\}_{k=1}^K$ as the relabeled set of asymptotic variance of $\{\sigma_k^2\}_{k=1}^K$ with an ascending order. For any allocation decisions $\bm{a}$ and $\tilde{\bm{a}}$ satisfying $\sum_{k=1}^i a_{(k)} \ge \sum_{k=1}^i \tilde{a}_{(k)}$ for $i=1,2,\cdots,K$, we have
{\setlength\abovedisplayskip{1.5pt}
\setlength\belowdisplayskip{1.5pt}
\begin{equation*}
\varsigma({\bm{a}}, \bm{w}^*({\bm{a}})) \le \varsigma({\tilde{\bm{a}}},\bm{w}^*({\tilde{\bm{a}}})).
\end{equation*}
}
In particular, the optimal allocation $\bm{a}^*$, which solves the optimization problem in Equation (\ref{eq:op2}), satisfies $a_k^* = \mathbbm{1}_{\{k=k^*\}}$ with the minimum asymptotic variance
{\setlength\abovedisplayskip{1.5pt}
\setlength\belowdisplayskip{1.5pt}
\begin{equation*}
\varsigma (\bm{a}^*,\bm{w}^*({\bm{a}^*})) = \sigma_{k^*}^2.
\end{equation*}
}
\end{theorem}

Theorem \ref{the:basic2} states that
an allocation decision $\bm{a}$ is more efficient, i.e., it induces a smaller $\varsigma(\bm{a},\bm{w}^*(\bm{a}))$, if it
allocates more budgets to more efficient statistics. In particular, if we greedily allocate all budgets to the most
efficient statistic $k^*$, the asymptotic variance $\varsigma(\bm{a},\bm{w}^*(\bm{a}))$ will be minimized.

Theorem \ref{the:basic1} and \ref{the:basic2} show that the optimal solutions are closely related to the asymptotic variances $\sigma_k^2$ of the individual statistics, and the directions for decreasing $\varsigma(\bm{a},\bm{w})$ are allocating as much budget to statistic $k^*$ as possible and weighting the individual estimators inversely proportional to their asymptotic variances. However, the asymptotic variances $\sigma_k^2$ are usually unknown before sampling. To address this challenge, we propose a two-stage framework, where we infer the best statistic $k^*$ in the first
stage before allocating all the remaining budget greedily in the second stage.

\section{Adaptive Two-Stage Framework}
\label{sec:pilot}
In this section, we first explain the basic concepts of a two-stage framework and then show the framework achieves higher sampling efficiency than two benchmark strategies, finally we propose an adaptive algorithm to determine an upper-bound of the optimal budget fraction which is allocated to the first stage.

\subsection{Two Benchmark Strategies and A Two-Stage Generalization}
\label{sec:basic}

Without knowing the asymptotic variances $\sigma_k^2$ of the individual statistics, we start with two naive strategies as benchmarks. The first strategy spends all budget $M$ on a randomly chosen statistic $k$;
the second strategy evenly divides the budget $M$ among $K$ statistics to construct the mixture estimator.
We call these two benchmark strategies as the {\em Random Statistics} (or RND) and {\em Average Statistics} (or AVG), respectively.

Based on the two benchmark strategies, we consider a two-stage generalization, which spends a partial budget to estimate the best statistic $k^*$ in a {\em pilot sampling stage} and allocates the remaining budget to an estimated best statistic $\hat{k}^*$ in a {\em regular sampling stage}.
We assume that a fraction $c\in[0,1]$ of the total budget $M$ is allocated for pilot sampling and name the $cM$ samples as the pilot budget. We evenly allocate the pilot budget among all $K$ statistics, and therefore, each statistic $k$ is allocated a budget of $m_k={cM}/{K}$ samples in this stage. We use these pilot samples to make an \emph{asymptotically unbiased estimate} of each asymptotic variance $\sigma_k^2$, and define the estimated value by $\hat{\sigma}_k^2(m_k)$.
Most likely, the statistic with the smallest estimated asymptotic variance tends to be the most efficient statistic $k^*$ for estimating $\bar{f}$. We call this statistic the \emph{inferred most efficient statistic} and denote it as $\hat{k}^*(cM)$, parameterized by the pilot sampling budget $cM$.
In the regular sampling stage, we allocate all the remaining sampling budget $(1-c) M$ to the inferred most efficient statistic $\hat{k}^*$, and fully use the total budget $M$ to construct a mixture estimator.

Under the above two-stage framework, we denote $\bm{a}(c)$ as the effective allocation decision, defined by
{\setlength\abovedisplayskip{1.5pt}
\setlength\belowdisplayskip{1.5pt}
\begin{equation}\label{eq:m1}
a_k(c) \triangleq {c}/{K} + (1-c)\cdot \mathbbm{1}_{\left\{k=\hat{k}^*(cM)\right\}},
\end{equation}
}
through which we can define the effective budget for each statistic $k$ as $m_k(c)\triangleq a_k(c)M$ naturally.
After both sampling stages, we construct a mixture estimator by using an estimated optimal weight vector $\hat{\bm{w}}^*(c)$.
We use the estimated value $\hat{\sigma}_k^2(m_k) $ to approximate $\sigma_k^2 $, and define $\hat{\bm{w}}^*(c)$ by substituting $\sigma_k^2$ with $\hat{\sigma}_k^2(m_k)$ in the optimal weight of Equation (\ref{eq:optiwei}) as
{\setlength\abovedisplayskip{1.5pt}
\setlength\belowdisplayskip{1.5pt}
\begin{equation}
\hat{w}_k^*(c) \triangleq \frac{a_k(c)}{\hat{\sigma}_k^{2}\left(m_k(c)\right)}/\displaystyle\sum_{i\in {\cal K}_a} \frac{a_i(c)}{\hat{\sigma}_i^{2}\left(m_i(c)\right)}, \quad \forall k\in {\cal K}_a.
\end{equation}
}
Consequently, the corresponding mixture estimator and its asymptotic variance can be written as $\hat{f}(\bm{a}(c),M,\hat{\bm{w}}^*(c))$ and $\varsigma(\bm{a}(c),\hat{\bm{w}}^*(c))$, respectively.

The two-stage framework actually uses the AVG and RND strategies in its pilot and regular sampling stages, respectively. In particular, the estimated statistic $\hat{k}^*$ plays the role of a random statistic in the RND strategy.
Also, the framework can be seen as a generalization of the two benchmark strategies, because the Average and Random Statistics are equivalent to a two-stage strategy of $c=1$ and $c=0$, respectively.


\begin{theorem}\label{the:bench}
The asymptotic variances of the Random Statistics and Average Statistics are $\varsigma(\bm{a}(0),\hat{\bm{w}}^*(0))$ and $\varsigma(\bm{a}(1),\bm{a}(1))$, respectively. They satisfy
{\setlength\abovedisplayskip{1.5pt}
\setlength\belowdisplayskip{1.5pt}
\begin{equation*}
\mathbb{E}\big[\varsigma(\bm{a}(0),\hat{\bm{w}}^*(0))\big] \!=\! \varsigma(\bm{a}(1),\bm{a}(1)) \!=\! \frac{1}{K}\displaystyle\sum_{k=1}^{K} \sigma_k^2.
\end{equation*}
}
\end{theorem}
{\setlength{\parindent}{0em}
Theorem \ref{the:bench} states that the expected asymptotic variance of the Random Statistics and the asymptotic variance of Average Statistics both equal the average of the asymptotic variances of all individual statistics.
}

\subsection{Asymptotic Performance of Two-Stage Strategies}
Our two-stage framework does not restrict how the asymptotic variances $\sigma_{k}^2$ are estimated in the pilot sampling stage. We will show that as long as $\hat{\sigma}_k^2(\cdot)$ is an asymptotically unbiased estimator for $\sigma_k^2$, the two-stage strategies will outperform the two benchmark strategies.
The detailed design of the estimator $\hat{\sigma}_k^2(\cdot)$ may depend on the sampling method of statistic $k$, and we will give an example of implementation in a later section.

Given any strategy $c\in[0,1]$, we can define the (unknown) optimal allocation decision as $\bm{a}^*(c)=(a^*_1(c),\dots,a^*_K(c))$ as
{\setlength\abovedisplayskip{1.5pt}
\setlength\belowdisplayskip{1.5pt}
\begin{equation*}
a^*_k(c) \!\triangleq  c/K +\big(1-c\big) \cdot\mathbbm{1}_{\left\{k=k^*\right\}}, \quad \forall k=1,\dots,K.
\end{equation*}
}
Under this optimal allocation $\bm{a}^*(c)$, by Theorem \ref{the:basic1}, the corresponding optimal weight vector becomes $\bm{w}^*(\bm{a}^*(c))$.
Intuitively, when a budget $cM$ is used to estimate each $\sigma^2_k$ in the pilot sampling stage, $m_k = cM/K$ for any statistic $k$ and the best statistic $k^*$ is more likely to induce a smaller estimated asymptotic variance $\hat{\sigma}_k^2(m_k)$ than other statistics. Consequently, the resulting allocation $\bm{a}(c)$ and weights $\hat{\bm{w}}^*(c)$ are more likely to be equal to the optimal $\bm{a}^*(c)$ and $\bm{w}^*(\bm{a}^*(c))$, respectively.
We consider the two-stage strategy $c$ as a function of the total budget $M$, denoted as $c(M)$, and simplify the notation $\bm{a}^*(c(M))$ as $\bm{a}^*(M)$. The next theorem shows that when the pilot budget fraction $c$ is higher than the order of $M^{-1}$, the two-stage strategy $c(M)$ is asymptotically optimal.

\begin{theorem}\label{the:cond}
Assume each estimated asymptotic variance $\hat{\sigma}_k^2(\cdot)$ is asymptotically unbiased for $\sigma_k^2$ $(k=1,\cdots,K)$, i.e.,
{\setlength\abovedisplayskip{1.5pt}
\setlength\belowdisplayskip{1.5pt}
\begin{equation*}
\hat{\sigma}_k^2(m_k) \xrightarrow{a.s.} \sigma_k^2 \quad as \quad m_k \rightarrow +\infty.
\end{equation*}
}
If $c(M) \in \omega\left(M^{-1}\right)$, i.e., for all $\delta >0$, there exists a positive number $M'$ such that
$c(M) \ge \delta M^{-1}$ for all $M > M'$,
{\setlength\abovedisplayskip{1.5pt}
\setlength\belowdisplayskip{1.5pt}
\begin{equation*}
\begin{split}
&\hat{k}^* \xrightarrow{a.s.} k^* \ , \ \bm{a}(c(M)) \xrightarrow{a.s.} \bm{a}^*(M) \ \ \text{and}\\
&\hat{\bm{w}}^*\big(c(M)\big) \xrightarrow{a.s.} \bm{w}^*\big(\bm{a}^*(M)\big) \ \ \text{as} \ \ M \rightarrow +\infty.
\end{split}
\end{equation*}
}
\end{theorem}

Theorem \ref{the:cond} shows that as the total budget $M$ grows, to guarantee an (asymptotic) optimal two-stage strategy, the fraction $c$ for the pilot budget does not need to be large. The condition $c(M) \in \omega(M^{-1})$ ensures that the pilot budget $c(M)M$ grows with $M$ unboundedly as $M$ goes to infinity, although $c$ itself could approaches zero, such that the estimated asymptotic variance $\hat{\sigma}_k^2(m_k)$ will converge to $\sigma_k^2$. Consequently, the two-stage strategy $c(M)$ will identify the most efficient statistic $k^*$ via the pilot sampling and set the optimal allocation $\bm{a}^*\big(c(M)\big)$ and optimal weight $\bm{w}^*\big(\bm{a}^*(c(M))\big)$ for the mixture estimator.

When we simply give the same weight for each sample point, for any allocation $\bm{a}$, the corresponding weight vector becomes $\bm{w}=\bm{a}$, which are proportional to their sample sizes.
To distinguish the benefit of choosing an optimal allocation $\bm{a}^*$ and an optimal weight $\bm{w}^*$, we consider an intermediate mixture estimator $\hat{f}\big(\bm{a}, M ,\bm{a}\big)$, which gets affected only by the allocation decision $\bm{a}$ and has an asymptotic variance $\varsigma\big(\bm{a},\bm{a}\big)$.

{\corollary \label{cor:1} Under the conditions of Theorem \ref{the:cond}, for any pilot fraction \(c(M) \in \omega(M^{-1})\), as $M \rightarrow +\infty$, we have
{\setlength\abovedisplayskip{1.5pt}
\setlength\belowdisplayskip{1.5pt}
\begin{equation*}
\begin{split}
&\varsigma\Big(\bm{a}\big(c(M)\big),\hat{\bm{w}}^*\big(c(M))\Big) \!\xrightarrow{a.s.} \varsigma\Big(\bm{a}^*\big(M\big),\bm{w}^*\big(\bm{a}^*(M)\big)\Big),
\end{split}
\end{equation*}
}
and the asymptotic limit of $\varsigma$ satisfies
{\setlength\abovedisplayskip{1.5pt}
\setlength\belowdisplayskip{1.5pt}
\begin{equation*}
\varsigma \left(\bm{a}^*(M),\bm{w}^*\left(\bm{a}^*(M)\right)\right) \le \varsigma \left(\bm{a}^*(M),\bm{a}^*(M)\right) \le \frac{1}{K}\displaystyle\sum_{k=1}^{K} \sigma_k^2
\end{equation*}
}
{\setlength\abovedisplayskip{1.5pt}
\setlength\belowdisplayskip{1.5pt}
\begin{equation*}
\text{and}\quad \varsigma\big(\bm{a}^*(M),\bm{w}^*(\bm{a}^*(M))\big) \le \frac{K \sigma_{k^*}^2}{K+(1-K)c(M)}.
\end{equation*}
}
}

As a consequence of Theorem \ref{the:cond}, Corollary \ref{cor:1} shows that as $M$ grows, the asymptotic variance $\varsigma$ induced by the strategy $c(M)$ converges to an optimal value $\varsigma \left(\bm{a}^*(M),\bm{w}^*\left(\bm{a}^*(M)\right)\right)$.
The first inequality implies that 1) using the estimated optimal weight $\hat{\bm{w}}^*(c(M))$ is more efficient than the equal weight $\bm{w}=\bm{a}$, and 2) using $\bm{w}=\bm{a}$ is again more efficient than the two benchmark strategies, whose (expected) asymptotic variances equal $\frac{1}{K}\sum_{k=1}^K\sigma_k^2$ as shown in Theorem \ref{the:bench}.
The second inequality provides an upper-bound for the optimal $\varsigma$, which can be derived from an estimator $\hat{f}_{\hat{k}^*}\big(a_{\hat{k}^*}(c(M)) M\big)$ which only uses the samples of the inferred best statistic $\hat{k}^*$ and throws out the samples of other statistics collected in the pilot sampling stage.

\subsection{Optimal Fraction for Pilot Budget}
Our design of any two-stage strategy $c(M)\in \omega(M^{-1})$ is asymptotically optimal. However, a more practical problem is that, given a finite budget $M$, how to choose an optimal fraction $c^*(M)$ for the pilot budget that maximizes the efficiency for the mixture estimator $\hat{f}$, i.e., $c^*(M)$ solves:
{\setlength\abovedisplayskip{1.5pt}
\setlength\belowdisplayskip{1.5pt}
\begin{equation}\label{eq:14cstar}
\begin{aligned}
\displaystyle \text{Minimize}  & \qquad Var\big(\hat{f}(\bm{a}(c),M,\hat{\bm{w}}^*(c))\big),\\
\quad\quad  \text{subject to}  & \quad \quad c\in [0,1].
\end{aligned}
\end{equation}
}
On the one hand, when allocating more budget for the pilot sampling, each $\hat{\sigma}_k^2(\cdot)$ could provide a more accurate estimation for the asymptotic variance $\sigma_k^2$ and the best statistic $k^*$ would have a higher chance to be picked out in the regular sampling stage.
On the other hand, increasing the pilot budget means that more budget will be allocated to some inefficient statistics at the pilot sampling stage. One needs to balance the above contradictory conditions so as to obtain an optimal fraction $c^*(M)$.
In practice, it is hard to obtain the exact value of the optimal fraction for the pilot budget $c^*(M)$, because it depends on the unknown values of asymptotic variances $\sigma_k^2$.
However, we will provide a heuristic algorithm to estimate $c^*(M)$ effectively, which is based on the following theoretical result on the monotonicity of $c^*(M)$.


\begin{theorem}\label{the:fraction}
Assume the rate of convergence of the estimated asymptotic variance $\hat{\sigma}^2_k(m)$ for each $\sigma^2_k$ is $\Theta(m^{-\eta_k})$, i.e., $\mathrm{sup} _{x\in \mathbb{R^+}}|G_{\hat{\sigma}^2_k(m)}(x) - G_{\sigma^2_k}(x)| = \Theta(m^{-\eta_k})$,
where $G_{\hat{\sigma}^2_k(m)}(x) = \mathbb{P}\left(\hat{\sigma}^2_k(m)\le x\right)$ and $G_{\sigma^2_k}(x) = \mathbbm{1}_{\{x\ge \sigma^2_k\}}$ are the cumulative distribution function of $\hat{\sigma}^2_k(m)$ and $\sigma^2_k$, respectively, and the order \(\eta_k >0\).
Let $\eta\!= \!\min\{\eta_k:k=1,\!\cdots\!,K\}$. The optimal fraction satisfies
\(\displaystyle\lim_{M\rightarrow +\infty} c^*(M) = 0\) with the rate of convergence \(\Theta(M^{-1+\frac{1}{\eta+1}})\).
\end{theorem}

Theorem \ref{the:fraction} shows that the optimal fraction $c^*(M)$ decreases to zero asymptotically with the rate $\Theta(M^{-1+\frac{1}{\eta+1}})$ when $M$ grows.
Intuitively, as the total budget $M$ increases, to guarantee the same accuracy for estimating ${k}^*$,
we only need to keep the pilot budget $cM$ constant and thus the fraction $c$ becomes smaller.
Both Theorem \ref{the:cond} and \ref{the:fraction} imply that when $M$ becomes larger, the optimal fraction $c^*(M)$ should decrease. Therefore, we assume that $c^*(M)$ follows a decreasing trend as $M$ increases (In Section \ref{sec:sim}, our evaluations in the Douban social network also support this conjecture well), based on which we propose an adaptive algorithm to dynamically determine the optimal fraction $c^*(M)$ for the pilot sampling.

\begin{algorithm}
\renewcommand{\algorithmicrequire}{\textbf{Input:}}
\renewcommand\algorithmicensure {\textbf{Output:} }
\caption{Adaptive Two-Stage Sampling $(M, \Delta_M)$}
\begin{algorithmic}[1]
\STATE $c \leftarrow \Delta_M/M$; \\
\STATE spend $\Delta_M$ budget for pilot sampling; \\
\STATE \textbf{while} $c < \hat{c}^*(cM)$ \textbf{do}\\
$\quad$ $c \leftarrow c + {\Delta_M}/{M}$; \\
$\quad$ spend $\Delta_M$ more budget for pilot sampling; \\
\textbf{end while}\\
\STATE choose the estimated best statistic $\hat{k}^*$;\\
\STATE spend the remaining budget $(1-c)M$ for regular sampling;\\
\end{algorithmic}
\label{alg:re}
\end{algorithm}

Algorithm \ref{alg:re} performs the pilot sampling in an adaptive manner. It takes two input parameters: the total budget $M$ and a budget spending stepsize $\Delta_M\in(0,M)$.
We denote $\hat{c}^*(\cdot)$ as a function where each $\hat{c}^*(m)$ provides an estimated upper-bound of the optimal fraction $c^*(M)$, when $m$ number of samples are used.
In step 3, we increase the pilot budget by $\Delta_M$ if the spent fraction $c$ is smaller than the derived upper-bound $\hat{c}^*(cM)$ for $c^*(M)$, until $c$ exceeds the upper-bound $\hat{c}^*(cM)$.
Based on the $cM$ samples generated in the pilot sampling stage, we choose the estimated best statistic $\hat{k}^*$ and spend the remaining budget $(1-c)M$ for regular sampling as usual.
In general, given any $m$ pilot samples, the function $\hat{c}^*(\cdot)$ uses them to estimate an optimal fraction $c^*(m')$ for some $m'< m$. Because $c^*(\cdot)$ has a decreasing trend in general, we could use this estimation of $c^*(m')$ as an upper-bound for $c^*(M)$ so as to determine whether the pilot sampling stage should end.
As the sampling budget $cM$ increases, the estimation $\hat{c}^*(cM)$ should decrease and approach $c^*(M)$, because it estimates some $c^*(m')$ and $m'$ increases.
Notice that our algorithm does not restrict how the upper-bound estimation $\hat{c}^*(\cdot)$ should be implemented, and we will provide an example of implementation which we use in our evaluation in a later section.
Finally, although a large stepsize approaches $c^*(M)$ faster, to avoid overestimating the pilot budget, a small value of $\Delta_M$ should be used in practice.

\section{Evaluation in Douban Social Network}\label{sec:sim}
In this section, we apply the adaptive two-stage framework to the Douban social network, a popular Chinese web site providing user comment and recommendation services for books, music and movies.
We first introduce multiple statistics which can be realized in Douban and then provide a detailed implementation of our framework to measure the statistics, finally we evaluate the performance of the framework.

\subsection{Multiple Statistics}

Similar to Twitter and Sina microblog, users in Douban can follow each other, and therefore, Douban can be seen as a followship graph\footnote{Here, we serve the followship graph as an undirected graph and one following relationship corresponds to an undirected edge.} in which the edges capture the following relationship. Douban also allows users create interest groups for others to join in. We consider two users who have a common group share a membership and Douban can also be seen as a membership graph. These two different social graphs, together with two random walk based sampling methods, the RWuR and FS introduced next,
provide four different available statistics.

\subsubsection{The random walk with uniform restarts (RWuR)}
The RWuR \cite{KAWAW} is a hybrid sampling method that mixes random walk crawling and uniform node sampling. It generates a sample set $\{X_t\}_{t\in \mathbb{N}}$ as follows. At each step $t$, assume the current node is $X_t \!= \!i$.
With probability \(\alpha/(d_{i}+\alpha)\), it jumps to an arbitrary node \(j\) of the graph chosen uniformly and make the transition $X_{t+1} \!= \!j$. With probability \(d_{i}/(d_{i}+\alpha)\), it uniformly chooses an \(i\)'s neighboring node $k$, i.e., $X_{t+1} \!= \!k$. The parameter \(\alpha\) (\(\ge 0\)) controls the probabilities of random walk and jump. Specially, when \(\alpha = 0\), the RWuR is the simple random walk, and when \(\alpha = +\infty\), the RWuR becomes the uniform node sampling.
Obviously, the sample set $\{X_t\}_{t\in \mathbb{N}}$ is biased towards the high-degree nodes. To correct the bias, it uses the Hansen-Hurwitz estimator \cite{33, 34} to re-weight the samples, i.e., the weight of the sample $X_t$ is inversely proportional to $d_{X_t}+\alpha$, and the unbiased estimator for $\bar{f}$ is
{\setlength\abovedisplayskip{1.5pt}
\setlength\belowdisplayskip{1.5pt}
\begin{equation}\label{eq:RWuR}
\hat{f}(m) = \displaystyle\sum_{t=1}^m \frac{f_{X_t}}{d_{X_t}+\alpha}/\displaystyle\sum_{t=1}^m \frac{1}{d_{X_t}+\alpha}.
\end{equation}
}

\subsubsection{The frontier sampling (FS)}
The FS \cite{ribeiro2010estimating} is a distributed sampling method that performs \(s\) (\(\in \mathbb{N}\)) random walkers on a graph. Initially, it uniformly obtains \(s\) nodes as the start nodes of the \(s\) random walkers. At each step, it first randomly selects the \(r\)-th walker with probability \(d_{v_r}/\sum_{i=1}^s d_{v_i}\), where \(v_i\) is the current node of the \(i\)-th walker. Then the \(r\)-th walker uniformly chooses a \(v_r\)'s neighboring node as the next sample and moves to it. Similar to the RWuR, the bias towards high degree nodes of the FS can be corrected by the Hansen-Hurwitz estimator, and the unbiased estimator for \(\bar{f}\) is
{\setlength\abovedisplayskip{1.5pt}
\setlength\belowdisplayskip{1.5pt}
\begin{equation}\label{eq:FS}
\hat{f}(m) = \displaystyle\sum_{t=1}^m \frac{f_{X_t}}{d_{X_t}}/\displaystyle\sum_{t=1}^m \frac{1}{d_{X_t}}.
\end{equation}
}

{The RWuR and FS samplers are less likely to get trapped in loosely connected components of a graph via jumping randomly and running multiple walkers, respectively. Thus, both of them usually perform better than the simple random walk with re-weighting \cite{A.H.Rasti2009Res, M.Gjoka2011Practical}, but we do not know which one achieves higher sampling efficiency in an unknown graph. Besides, it is unclear how the efficiencies of the two methods vary on the followship and membership graphs. Therefore, we choose the four statistics, the RWuR and FS on the followship and membership graphs, to demonstrate our two-stage framework.}

\subsection{Implementation of Two-Stage Framework}
Our adaptive two-stage strategy does not restrict how the estimators of the asymptotic variances \(\hat{\sigma}_k^2(\cdot)\) and the upper-bound estimation of the optimal fraction \(\hat{c}^*(\cdot)\) are constructed, as long as they are asymptotically unbiased. Next, we provide an example of detailed implementations of \(\hat{\sigma}_k^2(\cdot)\) and \(\hat{c}^*(\cdot)\) for measuring in Douban.

\subsubsection{Estimating the asymptotic variances}
\label{sec:imple2}
Both the pilot sampling and the adaptive Algorithm \ref{alg:re} need to estimate the unknown asymptotic variances $\sigma_k^2$.
Assume the sample set used to estimate $\sigma_k^2$ is collected by $q$ ($\ge 2$) samplers whose budgets are all $l$. We denote the estimated value for $\bar{f}$ based on the $j$-th sampler as $\hat{f}_k^{(j)}(l)$ for $j=1,\cdots,q$, which serves as a sample of the estimator $\hat{f}_k(l)$.
Then the sample variance of $\{\sqrt{l}\hat{f}_k^{(j)}(l):j=1,\cdots,q\}$ is defined by
{\setlength\abovedisplayskip{1.5pt}
\setlength\belowdisplayskip{1.5pt}
\begin{equation}\label{eq:sv}
S_k^2 (q,l) = \frac{l}{q-1} \displaystyle\sum_{j=1}^{q} \left[\hat{f}_k^{(j)}(l) - \frac{1}{q}\sum_{i=1}^{q} \hat{f}_k^{(i)}(l) \right]^2.
\end{equation}
}
It describes how far $\sqrt{l}\hat{f}_k^{(j)}(l)$ $(j=1,2,\cdots,q)$ are spread out. From the definition of asymptotic variance in Equation (\ref{eq:asy}), $S_k^2(q,l)$ is an asymptotically unbiased estimate of $\sigma_k^2$, i.e.,
{\setlength\abovedisplayskip{1.5pt}
\setlength\belowdisplayskip{1.5pt}
\begin{equation}
S_k^2 (q,l) \xrightarrow{a.s.} \displaystyle\lim_{l \rightarrow \infty} l  Var(\hat{f}_k(l)) =\sigma_k^2 \quad \text{as} \quad q,l \!\rightarrow\! \infty.
\label{eq:mav}
\end{equation}
}
Thus, we can use $S_k^2 (q,l)$ to estimate the asymptotic variance $\sigma_k^2$, i.e., \(\hat{\sigma}_k^2(ql) = S_k^2 (q,l)\). Also, because all unbiased graph sampling methods have the same definition of the asymptotic variance from Equation (\ref{eq:asy}), this implementation is applicable to any one of them. 

\begin{figure*}[!ht]
 \centering
 \subfigure[Number of followers]{
 \label{fig:1a}
 \includegraphics[width=0.386\textwidth]{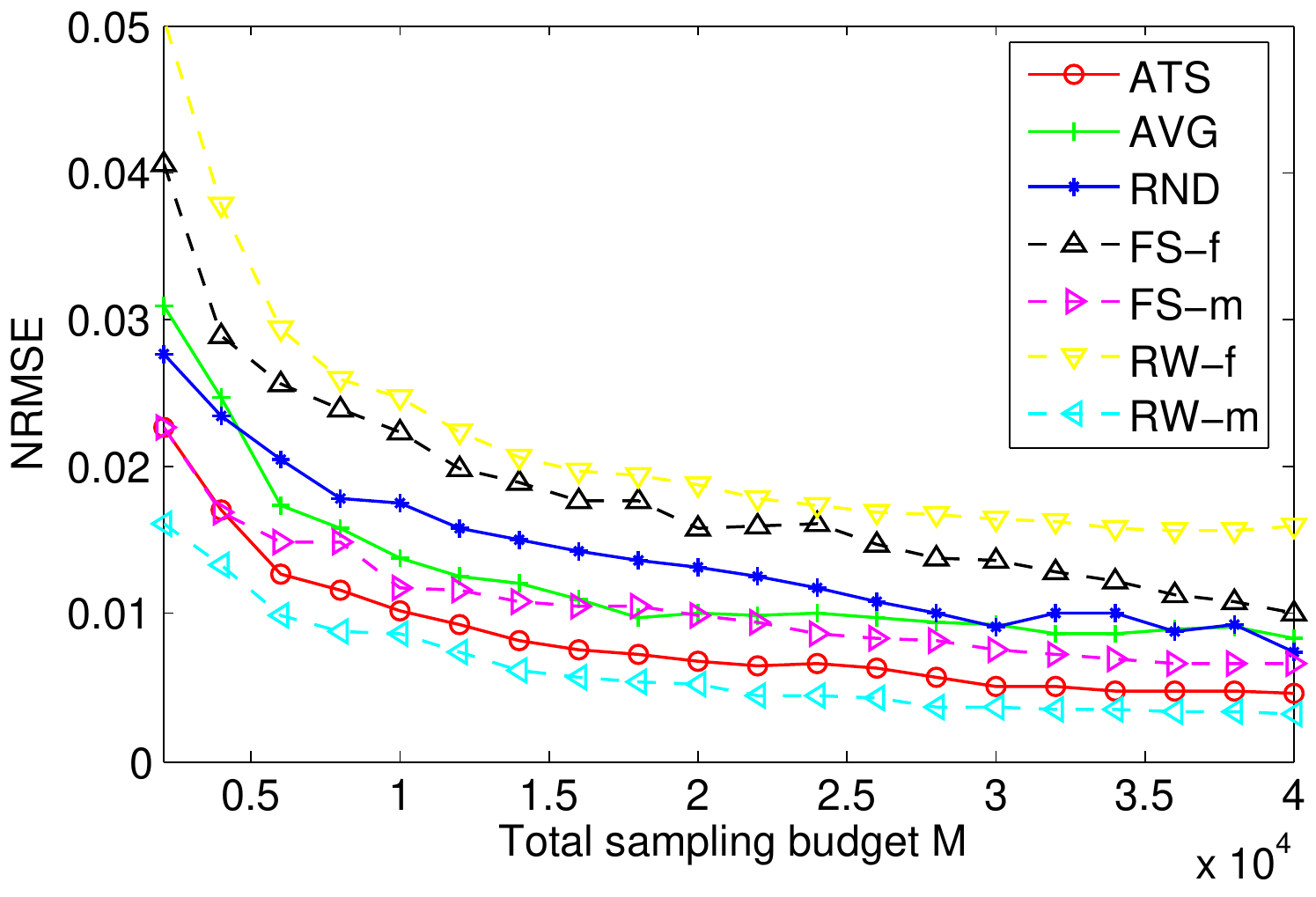}}
 \subfigure[Number of groups]{
 \label{fig:1b}
 \includegraphics[width=0.38\textwidth]{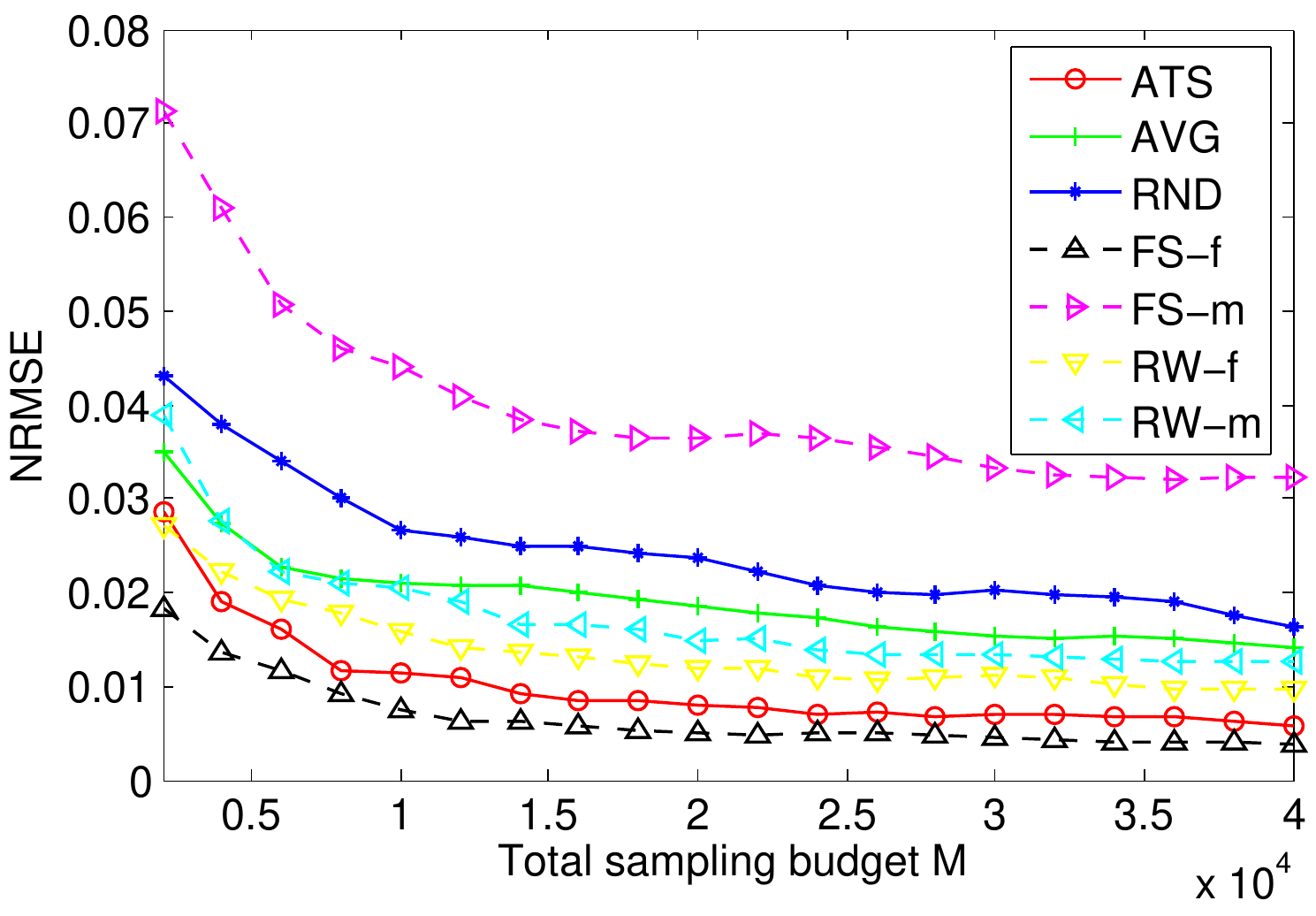}}
\caption{NRMSE of the adaptive two-stage strategy (ATS), Average Statistics (AVG), Random Statistics (RND), and individual statistics including the RWuR and FS methods on the followship graph (RW-f and FS-f) and on the membership graph (RW-m and FS-m), when we vary the total sampling budget $M$. (a) measure the average number of followers of users, and (b) measure the average number of interest groups of users.}
\vspace{-1em}
 \label{fig:1}
\end{figure*}
\subsubsection{Estimating upper-bound of optimal fraction $c^*(M)$}
Given any sub-budget $m<M$, we provide an implementation of the upper-bound estimation function $\hat{c}^*(m)$ as follows.
We use the budget $m$ to collect $B$ sample sets whose sizes are all $m' \triangleq \frac{m}{BK}$ for each statistic, and
by using these $m$ samples, we could estimate the optimal fraction $c^*(m')$ when the total given budget is $m'$.
We denote the $b$-th sample set of the $k$-th statistic as $S^{(b)}_k$. We could try different two-stage strategies with
fraction $c\in(0,1)$ on the $b$-th group of sample sets $\{S^{(b)}_k:k=1,\cdots,K\}$.
Specifically, like a normal two-stage strategy of fixed $c$, we obtain $\frac{cm'}{K}$ samples from each set $S^{(b)}_k$ as the pilot sampling, and use them to estimate the best statistic $k^*$, and then use the remaining $(1-c)m'$ samples of the inferred best statistic to generate a realization of the estimator $\hat{f}(\bm{a}(c),m',\hat{\bm{w}}^*(c))$. Finally, we calculate the sample variance of the $B$ realizations obtained from the $B$ groups of sample sets. Based on (\ref{eq:14cstar}), we choose the fraction $c$ that minimizes the sample variance as an estimation for $c^*(m')$, which serves as an upper-bound for the optimal fraction $c^*(M)$.

When increasing the number of realizations $B$, the estimation $\hat{c}^*(m')$ for $c^*(m')$ becomes more accurate. However, the budget $m' = \frac{m}{BK}$ decreases under a fixed budget $m$, and therefore, using $c^*(m')$ as an upper-bound for the optimal fraction $c^*(M)$ could be loose. As a result, we recommend to set the parameter $B$ moderately.

\subsection{Measurement Setup}
The publicly available information for every Douban user includes user-id, location, lists of followers, users he/she follows and the interest groups he/she joins in. We consider two measurement targets, i.e., the average number of followers of users and the average number of interest groups of users. To measure these targets, we develop crawlers to sample via the four statistics ($K=4$), i.e., the RWuR and FS methods on the followship and membership graphs. We ignore the users who do not have any followship or membership, as these isolated users cannot be visited via crawling.

We set the total sampling budget to be $M = 4\cdot 10^4$, which represents about $0.05\%$ of the total number of Douban users\footnote{By Nov. 15 2013, Douban service provider declare there are about 79.2 million users.}. For the statistics based on the FS method, we set the number of random walkers \(s=50\) in a FS sampler. For the statistics based on the RWuR method, we moderately set the parameter which controls the probabilities of random walk and jump \(\alpha = 0.1\). Besides, we also consider the cost of uniformly choosing a start node and jumping to an arbitrary node in the FS or RWuR method. This cost is about 14 units of budget in the Douban network, i.e., it needs to query an average of 14 randomly generated user-ids to obtain a valid one in the user-id space. In the two-stage framework, we set the number of realizations $B =10$ and the budget spending stepsize $\Delta_M = 2\%{M} = 800$ for Algorithm \ref{alg:re}. To estimate the asymptotic variances, we use $q = 5$ samplers.

We also implement the benchmark strategies, i.e., the Random Statistics and Average Statistics, and the four single-statistic strategies for comparison. 
To measure the estimation accuracy of the different sampling strategies, we use Normalized Root Mean Square Error (NRMSE) \cite{KAWAW, ribeiro2010estimating, M.Gjoka2011Practical}, $\sqrt{\mathbb{E}(\hat{f}-\bar{f})^2}/{\bar{f}}$
where $\bar{f}$ is the true value of the measurement target and $\hat{f}$ is the estimated one. Because the ``ground truth'' \(\bar{f}\) is not published by Douban, we calculate the NRMSE by taking as \(\bar{f}\) the grand average of \(\hat{f}\) values over all samples collected via all full-length crawlers and statistics. All experiment results presented in the following are the average of 25 independent simulations and our crawls were performed from Nov. 5th to 11th of 2013.

\subsection{Evaluation Results}
\subsubsection{Performance of the Adaptive Two-Stage Strategies (ATS)}

Figure \ref{fig:1} shows that the efficiencies of different statistics may vary for different measurement targets. For example, we observe that when we measure the average number of followers of users, the RWuR method on the membership graph (RW-m) leads to higher estimation accuracy than the FS method on the followship graph (FS-f) as shown in subfigure \ref{fig:1a}; however, when the target is the average number of interest groups of users, the conclusion is reversed as shown in subfigure \ref{fig:1b}. Thus, the efficiencies of the statistics vary as the measurement target changes and choosing a bad statistic, e.g., the RW-f strategy for estimating the average number of followers, may lead to an inaccurate estimation.
Without knowing the efficiencies of the individual statistics, Figure \ref{fig:1} shows that our adaptive two-stage strategy (ATS) always outperforms both benchmark strategies (AVG and RND) regardless of the measurement target.
Furthermore, our strategy (ATS) is only a bit inferior to the true best statistic (RW-m for estimating the average number of user's followers or FS-f for estimating the average number of user's groups), which could be used when the asymptotic variances of all statistics are known.
Figure \ref{fig:1} also demonstrates that our framework has good adaptivity for different measurement targets in the two subfigures.

\begin{figure}[t]
 \centering
 \subfigure[Number of followers]{
 \label{fig:2a}
 \includegraphics[width=0.233\textwidth]{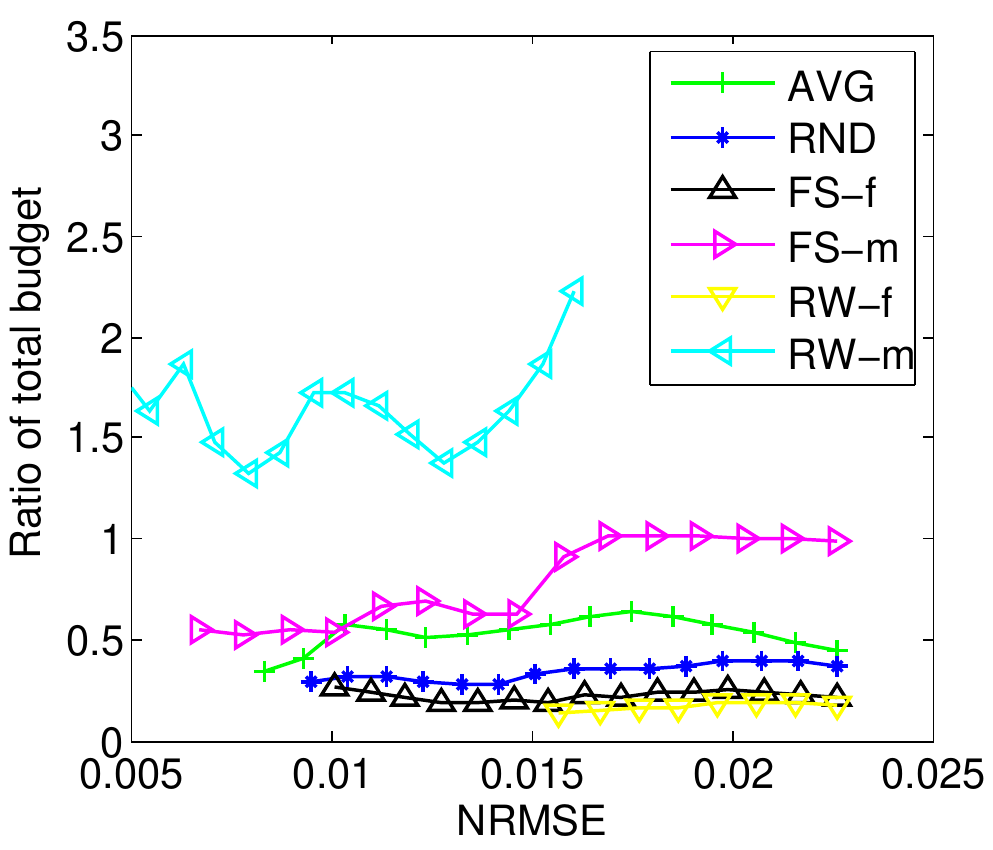}}
 \subfigure[Number of groups]{
 \label{fig:2b}
 \includegraphics[width=0.23\textwidth]{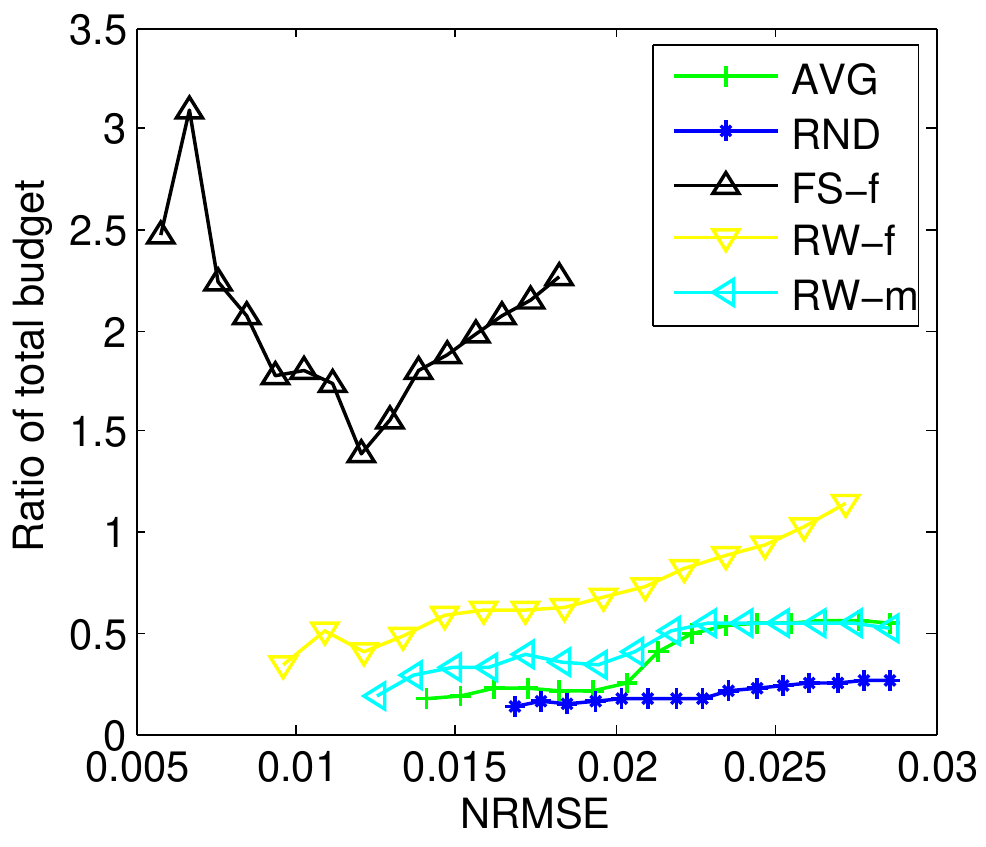}}
\caption{The ratio of needed budget between the two-stage strategy and others including the Average Statistics (AVG), Random Statistics (RND), and individual statistics including the RWuR and FS methods on the followship graph (RW-f and FS-f) and on the membership graph (RW-m and FS-m), so as to attain the same NRMSE. (a) measure the average number of followers of users, and (b) measure the average number of interest groups of users.}
 \label{fig:2}
 \vspace{-1em}
\end{figure}

\begin{figure}[t]
 \centering
 \subfigure[Number of followers]{
 \label{fig:3a}
 \includegraphics[width=0.227\textwidth]{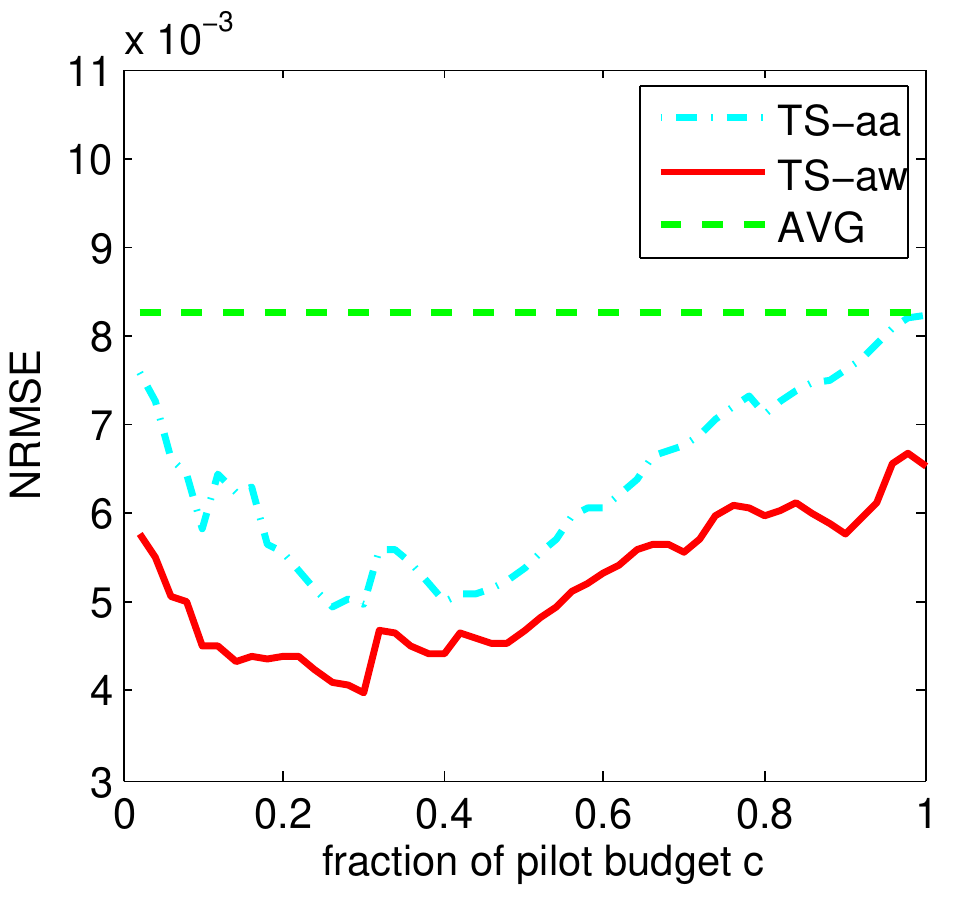}}
  \subfigure[Number of groups]{
 \label{fig:3b}
 \includegraphics[width=0.236\textwidth]{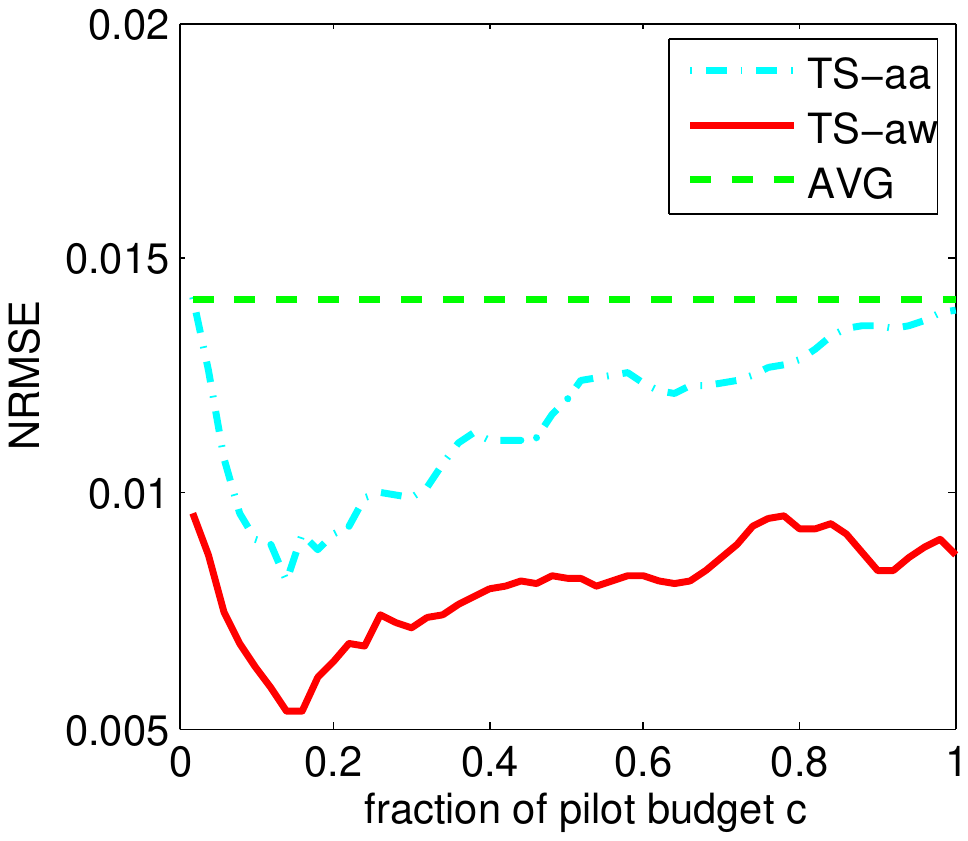}}
\caption{With total sampling budget $M=4\cdot 10^4$, NRMSE of the two-stage framework with the estimated optimal weights, i.e., $\hat{f}(\bm{a},M,\hat{\bm{w}}^*)$ (TS-aw), the two-stage framework with the same weight for each sample point, i.e., $\hat{f}(\bm{a},M,\bm{a})$ (TS-aa) and Average Statistics (AVG), when we vary the fraction of pilot budget $c$. (a) measure the average number of followers of users, and (b) measure the average number of interest groups of users.}
\vspace{-1em}
 \label{fig:3}
\end{figure}

Figure \ref{fig:2} shows the budget saving of our two-stage strategy (ATS) compared with the benchmarks (AVG and RND) and other single-statistic strategies if they can fulfill the given NRMSE target. For example, when measuring the average number of followers of users in subfigure \ref{fig:2a}, ATS saves about $49\%$ budget compared with the AVG strategy to obtain the $\text{NRMSE}\!=\!0.015$. From subfigures \ref{fig:2b}, as the target is the average number of groups of users, ATS saves about $75\%$ budget compared to the RND strategy for obtaining the $\text{NRMSE}=0.025$.
In general, we observe that our ATS strategy requires only $18\%$ to $57\%$ of the budget needed for the benchmark strategies to achieve the same NRMSE.
We also observe when measuring the average number of followers (resp. groups) of users, the best statistic RW-m (resp. FS-f) uses the smallest amount of budget, which is consistent with the observations from Figure \ref{fig:1} and the result of Theorem \ref{the:basic2}.

\subsubsection{Benefit of optimal allocation decision and weights }
The two-stage framework tries to improve estimation efficiency by choosing budget allocation decision and setting
estimated optimal weights for the mixture estimator. Figure \ref{fig:3} compares the NRMSE of our two-stage strategy when the weights are set to be equal (TS-aa) or optimally adjusted (TS-aw) and that of the AVG benchmark strategy, when the fraction $c$ of the pilot budget varies along the x-axis. We observe that the two-stage strategy with optimal weights always outperforms that with equal weights, which again outperforms the AVG benchmark strategy. Notice that under the equal weights, $c=0$ and $c=1$ corresponds to the RND and AVG strategies, respectively, which have the same performance as shown in Theorem \ref{the:bench}.
In general, when $c$ increases from $0$ to $1$, the benefit of two-stage strategy first increases and then decreases.
This is an integrated result of two competing factors: 1) increasing the pilot budget help select the more efficient statistic at the regular sampling stage, and 2) at the same time more budgets are allocated to the inefficient statistics at the pilot sampling stage. We also observe that the benefit of using optimal weights is larger when the pilot fraction is larger. The reason is that with the larger pilot budget, more samples are used on the inefficient statistics and therefore, optimal weights are more needed to discount those statistics.

\subsubsection{Effectiveness of the adaptive Algorithm \ref{alg:re}}\label{sec:alg1}
\begin{figure}[t]
 \centering
 \subfigure[Number of followers]{
 \label{fig:4a}
 \includegraphics[width=0.22\textwidth]{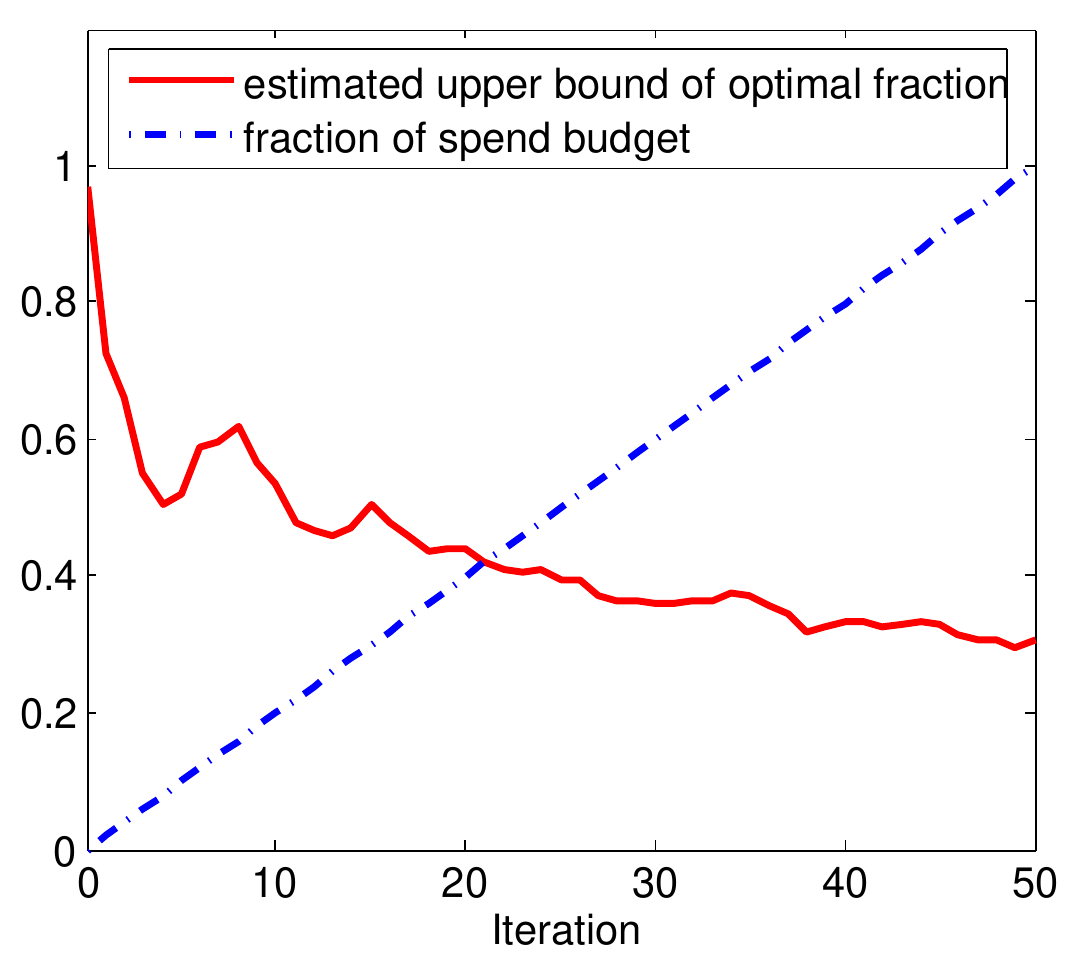}}
  \subfigure[Number of groups]{
 \label{fig:4b}
 \includegraphics[width=0.22\textwidth]{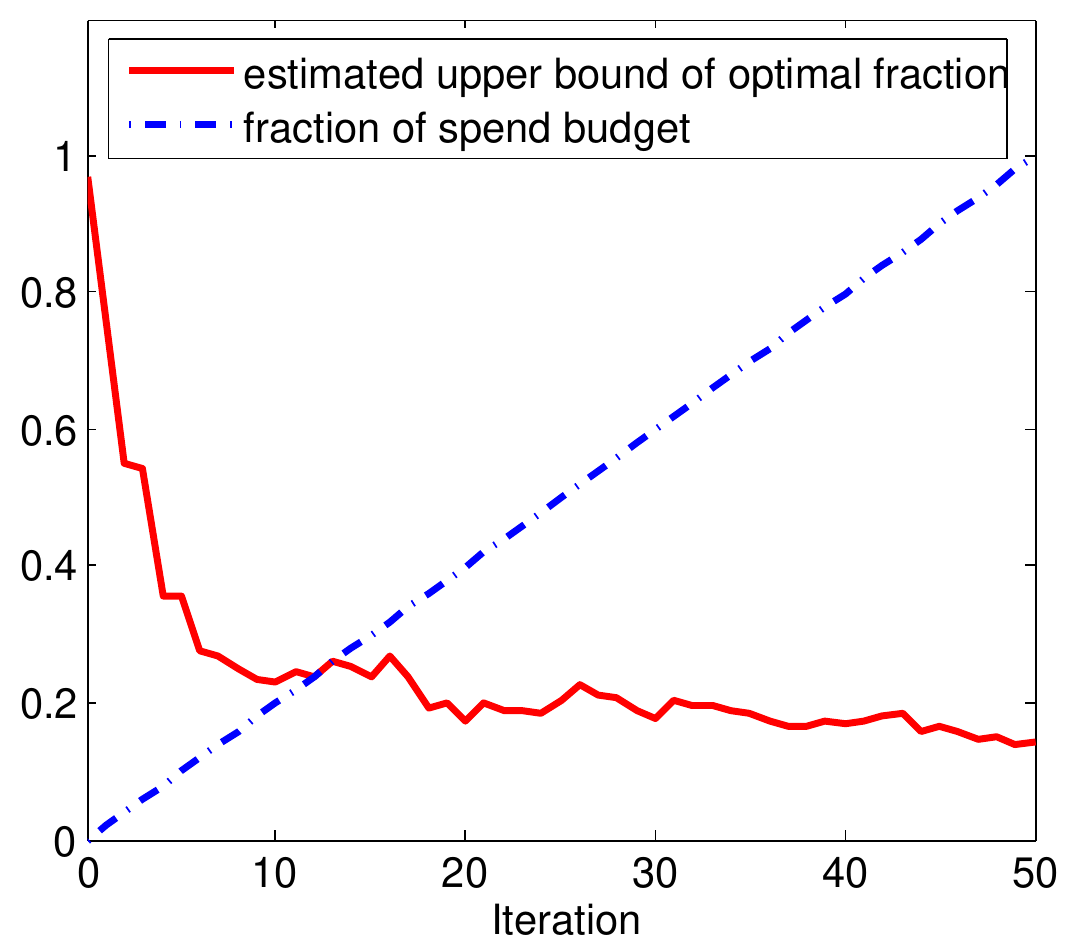}}
\caption{With total sampling budget $M=4*10^4$, the estimated upper bound of the optimal fraction $\hat{c}^*(cM)$ and the spent fraction of budget $c$ when the iteration increases in Algorithm \ref{alg:re}. (a) measure the average number of followers of users, and (b) measure the average number of interest groups of users.}
\vspace{-1.4em}
 \label{fig:4}
\end{figure}

We implemented Algorithm \ref{alg:re} for estimating the optimal pilot fraction. Figure \ref{fig:4} shows that the estimated optimal pilot fraction $\hat{c}^*(t\Delta_M)$ for ${c}^*(M)$ (solid line) has a decreasing trend as the number of iterations $t$ increases. It is consistent with our result that the optimal pilot fraction \(c^*(M)\) decreases as the budget \(M\) grows. The consumed fraction of the pilot budget $c$ (dash line) increases linearly (at a rate of $\Delta_M$) with the number of iterations. When the consumed pilot fraction $c$ is larger than the estimated upper-bound of optimal fraction, the iteration stops in Algorithm \ref{alg:re}. Subfigures \ref{fig:4a} (resp. \ref{fig:4b}) show that when measuring the average number of  users' followers (resp. groups), the estimated optimal pilot fraction $40\%$ (resp. $22\%$) approximates efficiently the real value $32\%$ (resp. $16\%$).
These results show that Algorithm \ref{alg:re} is effective for setting a near-optimal pilot fraction in the practical two-stage sampling.

\subsubsection{Observations of different statistics}
At last, we provide some insights into the different statistics. Subfigure \ref{fig:1a} indicates that the RWuR and FS methods on the membership graph perform better than them on the followship graph when we measure the average number of followers of users. However, when the target is the average number of groups of users, the conclusion is reversed as shown in subfigure \ref{fig:1b}. The reason may be that the followship (resp. membership) graph has a strong cluster feature \cite{schaeffer2007graph} that makes the samples highly correlated on the number of the users' followers (resp. groups). This strong correlation leads to a poor estimation accuracy.
We also observe that, for the followship graph, the FS method achieves higher efficiency than the RWuR; while the RWuR has smaller estimation error than the FS for the membership graph. Because the RWuR sampler frequently chooses an arbitrary node as restart on a less connected graph (e.g., the followship graph), which costs large budget and decreases the estimation accuracy. On the other hand, the RWuR is close to a single random walker on a well connected graph (e.g., the membership graph). Compared with the FS with multiple random walkers, it saves the cost of obtaining multiple uniform start nodes and converging to the walkers' steady state.

\section{Related Work}
\label{sec:related work}

\emph{Graph Sampling Techniques}. As OSN service providers rarely make publicly visible to the frame information of entire networks, most widely used graph sampling techniques in OSNs are crawling methods. Early graph crawling methods are based on Breath-First Search (BFS), Depth-First Search (DFS) and Snowball Sampling (SBS) \cite{DD}. In particular, BFS has been frequently used to explore large networks, such as Youtube and Facebook \cite{33}. However, these methods introduce a large bias towards high degree nodes and it is difficult to be corrected in general graphs \cite{2, 8, 18, 26}.

Recently the most popular graph crawling is random walk-based sampling, including simple random walk with re-weighting (RWRW) \cite{A.H.Rasti2009Res, M.Gjoka2011Practical} and Metropolis-Hastings random walk (MHRW) \cite{W.K.Hastings1970Monte}. RWRW is considered as a special case of Respondent-Driven Sampling (RDS) \cite{A.H.Rasti2009Res} if only one neighbor is chosen in each iteration and revisiting nodes is allowed. It is also biased to sample high degree nodes, but the bias can be corrected by the Hansen-Hurwitz estimator shown in \cite{33, 34}. RWRW was not only used to sample OSNs \cite{34, 18}, but also P2P networks and Web \cite{22,37}. MHRW is based on the Metropolis-Hastings (MH) algorithm and provides unbiased samples directly \cite{W.K.Hastings1970Monte, M.Gjoka2011Practical}. Some studies \cite{A.H.Rasti2009Res, M.Gjoka2011Practical} have shown that RWRW estimates are more accurate than MHRW estimates.

\emph{Improvement of sampling efficiency}. Researchers have proposed some methods to improve the sampling efficiency against random walk-based sampling, including the FS \cite{ribeiro2010estimating} and RWuR \cite{KAWAW} methods which we apply as showcases in this work. Besides, Kurant et al. \cite{kurant11_magnifying} presented a weighted random walk method to perform stratified sampling with a priori estimate of network information. Lee et al. \cite{lee2012beyond} proposed a non-backtracking random walk which forbids the sampler to backtrack to the previously visited node, and they theoretically guaranteed the technique achieves higher efficiency than a simple random walk. Our work concentrates on how to combine the existing statistics (sampling methods) efficiently and thus is complementary to their approaches.

It is worth mentioning that, Gjoka et al.\cite{start} designed a multi-graph sampling technique for the social networks which have multiple relation graphs. Their technique improves the convergence rate of the sampler by walking along a union graph of all relations. But it does not distinguish the efficiencies of walking on different relation graphs. In this paper, we propose the two-stage framework to select an inferred most efficient one from multiple graphs to improve sampling efficiency further.

\section{Conclusions}
\label{sec:conclusion}

In this paper, we consider the problem of using multiple statistics to efficiently sample online social networks.
Given a fixed sampling budget, we design budget allocation decisions and combine them to construct an optimal estimator.
In particular, we formulate a mixture sampling problem which constructs the optimal mixture estimator, and derive the optimal weights and a condition of ranking budget allocation decisions for the optimal estimator.
Because the asymptotic variances of the individual statistics are unknown in practice, we propose an adaptive two-stage framework, 
which spends a partial budget to test all different statistics in the pilot sampling stage and allocates the remaining budget to the inferred best statistic in the regular sampling stage.
To optimally set the sub-budget for the pilot sampling stage, we design an adaptive algorithm to dynamically decide an upper-bound of the optimal pilot budget and test whether the pilot sampling should end.
We implement the adaptive two-stage framework and evaluate its performance in the Douban network. We demonstrate, in theory and experiment, that our two-stage framework achieves higher sampling efficiency than two benchmark strategies.

\nocite{*} 

\begin{footnotesize}
\bibliographystyle{IEEEtran}
\bibliography{tex}
\end{footnotesize}

\appendix
\noindent
\textbf{Proof of Theorem \ref{the:basic1}:} From Equation (\ref{eq:ap_1}), we have
\begin{equation*}
\begin{split}
\lim_{M\rightarrow \infty} &\hat{f}(\bm{a},M,\bm{w}) = \lim_{M\rightarrow \infty} \sum_{k\in {\mathcal K}_a} w_k \cdot \hat{f}_k (a_kM)\\
&= \sum_{k\in {\mathcal K}_a} w_k \cdot \lim_{M\rightarrow \infty}  \hat{f}_k (a_kM) \xrightarrow{a.s.} \sum_{k\in {\mathcal K}_a} w_k \bar{f}
\end{split}
\end{equation*}
implying that the mixture estimator \(\hat{f}(\bm{a},M,\bm{w})\) is asymptotically unbiased for \(\bar{f}\) if and only if \(\sum_{k\in {\mathcal K}_a} w_k = 1\), i.e., Equation (\ref{eq:ap_2}) concludes. Then from Equation (\ref{eq:14asy}), observe that
\begin{equation*}
\begin{split}
\varsigma&(\bm{a},\bm{w}) = \displaystyle\lim_{M \rightarrow \infty} M  \cdot Var(\hat{f}(\bm{a},M,\bm{w}))\\
&= \displaystyle\lim_{M \rightarrow \infty} M \sum_{k \in {\mathcal K}_a} w_k^2 \cdot Var(\hat{f}_k(a_kM))\\
&= \sum_{k \in {\mathcal K}_a} \frac{w_k^2}{a_k} \displaystyle\lim_{M \rightarrow \infty} a_kM\cdot Var(\hat{f}_k(a_kM))= \displaystyle\sum_{k\in {\mathcal K}_a} \!\frac{w_k^2}{a_k} \cdot\sigma_k^2.
\end{split}
\end{equation*}
Based on Cauchy-Schwarz inequality, it satisfies
\begin{equation*}
\left[\displaystyle\sum_{k\in {\mathcal K}_a} {w_k}^2\cdot\frac{\sigma_k^{2}  }{a_k}\right]\cdot
\left[\displaystyle\sum_{k\in {\mathcal K}_a} \frac{a_k}{\sigma_k^{2} } \right]
\ge\left[\displaystyle\sum_{k\in {\mathcal K}_a} {w_k}\right]^2 = 1
\end{equation*}
where the equality holds up if and only if $
w_k = \frac{a_k}{\sigma_k^{2}}/\sum_{i\in {\mathcal K}_a} \frac{a_i}{\sigma_i^{2}}.
$
Thus given an allocation decision \(\bm{a}\), for any weight vector \(\bm{w} \in {\cal W}_{\bm{a}}\),
\begin{equation*}
\begin{split}
\varsigma_{\bm{a}}(\bm{w}) \!= \!\displaystyle\sum_{k\in{\mathcal K}_a} {w_k}^2\!\cdot\!\frac{\sigma_k^{2}}{a_k}
\!\ge\! \left[\displaystyle\sum_{k\in{\mathcal K}_a} \frac{a_k}{\sigma_k^{2}} \right]^{-1}\!\!\!\! \!=\! \varsigma_{\bm{a}}(\bm{w}^*)
\end{split}
\end{equation*}
holds up, i.e., \(\bm{w}^*\) solve the optimization problem in Equation (\ref{eq:op1}).\\
$\,$\\
\textbf{Proof of Theorem \ref{the:basic2}:} If the allocation decisions $\bm{a}$ and $\bm{a}'$ satisfies \(\sum_{k=1}^i a_{(k)}\ge \sum_{k=1}^{i} a'_{(k)}\) (\(i\!=\!1,\!\cdots\!,K\)),
\begin{equation*}
\begin{split}
\displaystyle\sum_{k=1}^K\! \frac{a_{(k)} \!-\! a'_{(k)}}{\sigma_{(k)}^2}\! &\ge \! \frac{a_{(1)}\!+\!a_{(2)}\!-\!a'_{(1)}\!-\!a'_{(2)}}{\sigma_{(2)}^2} + \!\displaystyle\sum_{k=3}^K \!\frac{a_{(k)} \!-\! a'_{(k)}}{\sigma_{(k)}^2}\\
&\ge \cdots \ge\frac{\sum_{k=1}^K [a_{(k)}-a'_{(k)}]}{\sigma_{(K)}^2} = 0.
\end{split}
\end{equation*}
holds up. Based on Theorem \ref{the:basic1}, we have
\begin{equation*}
\varsigma({\bm{a}}, \bm{w}^*({\bm{a}})) \!= \!\left[\displaystyle\sum_{k\in {\mathcal K}_a} \frac{a_{(k)}}{\sigma_{(k)}^{2}}\right]^{-1} \!\!\!\!\!\!\le \left[\displaystyle\sum_{k \in {\mathcal K}_a} \frac{a'_{(k)}}{\sigma_{(k)}^{2}}\right]^{-1} \!\!\!\!\!\!\!= \varsigma({\bm{a}'},\bm{w}^*({\bm{a}'})).
\end{equation*}
In particular, for any \(\bm{a}\), the allocation \(\bm{a}^*\) satisfies
$
\varsigma({\bm{a}}, \bm{w}^*({\bm{a}})) \ge \varsigma({\bm{a}^*}, \bm{w}^*({\bm{a}}^*)) = \sigma_{k^*}^2.
$
\\
\\
\textbf{Proof of Theorem \ref{the:bench}:} When \(c=1\), \(a_k(1) = 1/K\) (\(k=1,\!\cdots\!,K\)). Then we have $\varsigma(\bm{a}(1),\bm{a}(1)) = \frac{1}{K} \sum_{k=1}^K \sigma_k^2 $ from Equation (\ref{eq:ap_3}).
When \(c=0\), the inferred most efficient statistic is uniform randomly chosen, i.e., \(P(\hat{k}^*(cM) = k) =1/K\) (\(k=1,2,\cdots,K\)). Then \(P (a_k(0) = 1) = 1/K\) and
$
\mathbb{E}(\varsigma(\bm{a}(0),\hat{{\bm{w}}^*}(0))) = \sum_{k=1}^K P(a_k(0) = 1)\cdot \sigma_k^2= \frac{1}{K} \sum_{k=1}^K \sigma_k^2.
$
\\
\\
\textbf{Proof of Theorem \ref{the:cond}:} As each estimated asymptotic variance \(\hat{\sigma}_k^2(\cdot)\) is an asymptotically unbiased for \(\sigma_k^2\) \((k=1,\cdots,K)\), observe that
\begin{equation*}
\begin{split}
&\lim_{M \rightarrow \infty} P(\hat{k}^* = k^*) \\
= &\lim_{M \rightarrow \infty} P(\hat{\sigma}_{k^*}^2(\frac{c(M)M}{K}) \le \hat{\sigma}_j^2(\frac{c(M)M}{K}) \quad\forall j=1,\cdots,K)\\
=&\lim_{M \rightarrow \infty} P(\sigma_{k^*}^2 \le \sigma_j^2 \quad \forall j=1,\cdots,K) = 1
\end{split}
\end{equation*}
holds up if \(c(M)\in\omega (M^{-1})\). Thus, we have, as \(M \rightarrow \infty\),
\begin{equation*}
\begin{split}
a_k(c(M)) \xrightarrow{a.s.} \frac{c(M)}{K} + (1-c(M))\cdot \bm{1}_{\{k=k^*\}} = a^*_k(M)
\end{split}
\end{equation*}
for $\forall k=1,\cdots,K.$ Consequently, it satisfies
$\hat{k}^* \xrightarrow{a.s.} k^* $,$\bm{a}(c(M)) \xrightarrow{a.s.} \bm{a}^*(M)$ and
$\hat{\bm{w}}(c(M)) \xrightarrow{a.s.} \bm{w}^*(\bm{a}^*(M))$ as $M \rightarrow +\infty$.
\\
\\
\textbf{Proof of Corollary \ref{cor:1}:} From Theorem \ref{the:basic1}, for any $c(M)\in w(M^{-1})$, we have $\varsigma(\bm{a}^*(M),\bm{w}^*(\bm{a}^*(M))) \le \varsigma\left(\bm{a}^*(M),\bm{a}^*(M)\right)$ as $M\rightarrow \infty$, where
\begin{equation*}
\begin{split}
&\varsigma  (\bm{a}^*(M),\bm{w}^*(\bm{a}^*(M))) = \left[\displaystyle\sum_{k=1}^K \frac{a_k^*(M)}{\sigma_k^2}\right]^{-1}\\
&\le \left[\frac{a_{k^*}^*(M)}{\sigma_{k^*}^2}\right]^{-1} =  \frac{K\sigma_{k^*}^2}{K+(1-K)c(M)},\\
&\varsigma(\bm{a}^*(M),\bm{a}^*(M))= (1\!-\!c(M))\sigma_{k^*}^2 \!+\!  \frac{c(M)}{K} \\
&=\frac{1}{K}\displaystyle\sum_{k=1}^K \sigma_k^2 -\frac{1-c(M)}{K}\displaystyle\sum_{k=1}^K \big(\sigma_k^2-\sigma_{k^*}^2\big) \le \frac{1}{K}\displaystyle\sum_{k=1}^{K} \sigma_k^2.
\end{split}
\end{equation*}
\\
\\
\textbf{Proof of Theorem \ref{the:fraction}:} Because $\mathbb{E}\big[\varsigma(\bm{a}(c(M)),\hat{\bm{w}}^*(c(M)))\big] = \displaystyle\lim_{M\rightarrow +\infty} M\cdot Var\big(\hat{f}(\bm{a}(c),M,\hat{\bm{w}}^*(c))\big)$ from Equation (\ref{eq:14asy}), the fraction \(c^*(M)\) minimizes  $\mathbb{E}\big[\varsigma(\bm{a}(c(M)),\hat{\bm{w}}^*(c(M)))\big]$.
When the convergence rate of estimated asymptotic variance $\hat{\sigma}^2_k(m)$ for $\sigma^2_k$ is $\Theta(m^{-\eta_k})$ $(k=1,\cdots,K)$ and \(c(M)\in\omega (M^{-1})\), it satisfies $\mathbb{E}\big[\varsigma(\bm{a}(c(M)),\hat{\bm{w}}^*(c(M)))\big] \rightarrow \mathbb{E}\big[\varsigma(\bm{a}^*(M),\bm{w}^*(\bm{a}^*(M)))\big] \ \ \text{as} \ \ M \rightarrow +\infty$ with the convergence rate $\Theta\big((c(M)M)^{-\eta}\big)$ from Theorem \ref{the:cond} and Bounded Convergence Theorem \cite{titchmarsh1939theory}.
Further, as \(c^*(M)\) minimizes $\mathbb{E}\big[\varsigma(\bm{a}(c(M)),\hat{\bm{w}}^*(c(M)))\big]$, it satisfies the first-order condition $\displaystyle\lim_{M\rightarrow +\infty}\frac{d\mathbb{E}\big[\varsigma(\bm{a}(c(M)),\hat{\bm{w}}^*(c(M)))\big]}{dc(M)}\big|_{c = c^*(M)} = 0$, and therefore $\Theta(\frac{d(cM)^{-\eta}}{dc}\big|_{c = c^*(M)}) = \Theta\big(\!\!-\eta M^{-\eta} {c^*(M)}^{-\eta-1}\!\big) = \Theta (\frac{d\mathbb{E}\big[\varsigma(\bm{a}^*(M),\bm{w}^*(\bm{a}^*(M)))\big]}{dc}\big|_{c=c^*(M)}) = \Theta(1)$, from which we can derive that $c^*(M) = \Theta( M^{\frac{-\eta}{\eta + 1}}) = \Theta(M^{-1+\frac{1}{\eta + 1}})$ and $\displaystyle\lim_{M\rightarrow +\infty} c^*(M) = 0$. 
\end{document}